# COMPARATIVE ANALYSIS OF GALLEX AND GNO SOLAR NEUTRINO DATA

P.A. Sturrock[1], D.O. Caldwell[2], and J. D. Scargle[3]

ADDRESSES

[1] Center for Space Science and Astrophysics, Varian 302, Stanford University, Stanford, California 94305-4060

[2] Physics Department, University of California, Santa Barbara, CA 93106-9530

[3] NASA/Ames Research Center, MS 245-3, Moffett Field, CA 94035

ABSTRACT

Since the GALLEX and GNO datasets were derived from closely related experiments, there is a natural tendency to merge them. This is perhaps appropriate for any analysis based on the hypothesis that the solar neutrino flux is constant, but it is not necessarily appropriate for an analysis that allows for possible variability, since the GALLEX and GNO experiments belong to different solar cycles. Moreover, we find significant differences between the GALLEX and GNO datasets. It appears, from inspection of the time series and histograms, that GNO measurements are compatible with the assumption that the solar neutrino flux is constant, but GALLEX measurements are not. Furthermore, power-spectrum analysis yields evidence of rotational modulation in GALLEX data but not in GNO data. We compare our results with those of Pandola, who claims that GALLEX-GNO data show no evidence for variability.

# 1. INTRODUCTION

In earlier articles [1-7], we have presented evidence for variability of the solar neutrino flux as measured by the Homestake [8], GALLEX [9 - 13] and GNO [14 - 16] experiments. We have also presented strong evidence of variability [17 - 20] in data acquired by the Super-Kamiokande experiment [21, 22], and responded to Superkamiokande publications [23 - 25] that dispute the evidence for variability. Concerning GALLEX-GNO data, Pandola [26] has recently presented an analysis of the combined dataset, which he claims "does not support the presence of a time variability with characteristic periods resembling those of rotation of the solar magnetic fields…." The GNO Collaboration [16] recently stated that "the distribution of the individual run results is consistent with the hypothesis of a neutrino flux that is constant in time," but they also state that this "does not invalidate other hypotheses that might give similar or even better [short] time dependent fits."

The present article provides a comparative analysis of GALLEX data [13] and recently released GNO data [16]. Section 2 contains a discussion of the neutrino and solar physics that could in combination lead to variability of the solar neutrino flux, providing a physical reason why the GALLEX and GNO experiments, which were carried out in different solar cycles, should be analyzed separately. Section 3 presents a preliminary comparison (before embarking on power spectrum analysis) of GALLEX and GNO data. We examine the histograms formed from flux estimates and from upper and lower error estimates of the two datasets and find that they differ quite significantly. (The difference between the GALLEX and GNO error estimates is found to be a 5-sigma effect.) We next examine time variations by forming running means of the GALLEX and GNO flux estimates, and find that these also differ significantly. In addition, we note a difference between maximum-likelihood estimates of the flux (assumed constant) formed from the two datasets: when we divide each dataset into five parts we find that, according to the chi-square test, the GNO dataset is compatible with a constant neutrino flux, but the GALLEX dataset is not.

In Section 4, we carry out some preliminary examinations of the combined GALLEX-GNO dataset. We examine the periodogram formed from the end-times, and find a huge peak at 52.18 yr$^{-1}$, due to the fact that run durations are typically multiples of one week. This periodicity must be expected to lead to aliasing of any power spectrum. We also carry out a power-spectrum analysis of the combined dataset, using the Lomb-Scargle procedure, assigning flux measurements to the end times of runs, and find that the resulting power spectrum is quite similar to one obtained by Pandola [26].

In Section 5, we analyze the GALLEX, GNO, and combined datasets by an extension of the Lomb-Scargle procedure that takes account of both the start time and end time of each run. The resulting power spectrum for the combined dataset is similar, but not identical, to that obtained by the basic Lomb-Scargle method in Section 4. We find quite strong peaks in a rotational search band in the power spectrum derived from GALLEX data, but not in that derived from GNO data.

In Section 6, we carry out significance estimates of the principal peaks in the GALLEX power spectrum by a Monte Carlo procedure in which we apply an identical power spectrum analysis to surrogate datasets that have statistical properties identical to that of the actual dataset. Our findings are discussed in Section 7. For reasons explained in that section, power-spectrum analysis that takes account of the error estimates is deferred for a later article. In Appendix A, we discuss evidence in GALLEX data for modulation at a harmonic of the rotational frequency. In Appendix B, we point out that some of the peaks in the power spectra appear to be related to frequencies that we expect from r-mode oscillations.

2. PHYSICS REPONSIBLE FOR SOLAR NEUTRINO FLUX MODULATION

We have presented a particle physics explanation for solar neutrino flux modulation that is compatible with all solar data (even improving fits to the time-averaged data) and is compatible with all known constraints [7]. This model, which has recently been analyzed in detail by Chauhan and Pulido [27], is here reviewed very briefly.

The needed process for flux modulation is Resonant-Spin-Flavor-Precession [28 - 30] (RSFP), but only as an effect on the neutrino flux subdominant to matter-enhanced neutrino oscillations, specifically the favored LMA solution [31]. Since the KamLAND experiment [32] (which observes oscillations in no magnetic field) gives parameters compatible with LMA for $v_e$ oscillations in a magnetic field, RSFP cannot be the sole source of the solar neutrino deficit. To fit solar neutrino data, however, RSFP requires a neutrino mass-square difference between the lower mass state of the $v_e$ and the final state which is very different $\left(\Delta m^2 \sim 10^{-8} eV^2\right)$ from that which LMA requires. Since the three active light neutrinos provide only two mass differences, whereas a third is needed to meet this requirement, a new, sterile neutrino is necessary. This particle does not need to mix with active neutrinos, but is coupled to the electron neutrino only via a transition magnetic moment, as is necessary for RSFP. This lack of mixing also avoids all constraints on sterile neutrinos. Because of the mass difference involved, the RSFP effects occur in the solar convection zone, whereas the LMA oscillation takes place at a much smaller solar radius. In consequence, the RSFP process occurs in a region of strong magnetic field. It is important to note that the magnetic field in the convection zone changes with the solar cycle, and this in turn affects the resonant frequencies.

GALLEX data acquisition was confined to solar cycle 23, whereas GNO data acquisition was confined to solar cycle 24. Since each solar cycle is unique, we must infer that the magnetic-field structure of each cycle is unique. Since solar neutrino flux modulation depends on the magnetic field, it follows that it is essential to analyze the two experiments separately.

## 3. COMPARISON OF GALLEX AND GNO DATA

We have noted that there is a physical reason for treating GALLEX and GNO datasets separately, namely that the experiments belong to different solar cycles. In this section, we find that there are also statistical reasons to treat them separately.

At a time when only a few (19) measurements were available from the GNO experiment, we found that the combined GALLEX-GNO flux measurements had a bimodal distribution [4]. However, 58 measurements are now available for GNO, comparable with the 65 measurements of the GALLEX experiment, so we now have enough data to compare separate histograms formed from the GALLEX and GNO datasets. We show in Figure 1 histograms formed from the experimental flux estimates $g_r$, where $r$ enumerates the runs, derived from the GALLEX and GNO experiments and from the combined dataset. The GALLEX histogram clearly appears to be bimodal, with a dip in the range 60 to 100 SNU. The structure of the GNO histogram appears quite different from the GALLEX histogram, and is not so obviously bimodal, although one finds that there is a dip in the range 70 – 90 SNU. Furthermore, when we combine the GALLEX and GNO data, the flux histogram appears to be bimodal.

We show in Figure 2 plots of the upper and lower error estimates $\sigma_{u,r}$ and $\sigma_{l,r}$, respectively, from the GALLEX and GNO experiments. We see that both sets of error estimates are clearly asymmetric, and that the GNO error estimates are appreciably smaller than the GALLEX error estimates. We present a quantitative comparison of the flux and error measurements for the two experiments in Table 1. The standard error of the mean is given by

$$sem(g) = \frac{std(g)}{sqrt(N(g))}, \qquad (3.1)$$

where $std(g)$ is the standard deviation and $N(g)$ is the number of data points. The entry "sigma" is the root-mean-square of the GALLEX and GNO standard-error-of-the-mean terms.

The difference between the GALLEX and GNO flux measurements is a 1.9 sigma effect, that can occur by chance with (two-tail estimate) probability 0.05. However, the difference between the GALLEX and GNO upper error measurements is a 5.4 sigma effect, which can only occur by chance with probability $4 \times 10^{-8}$, and the difference between the GALLEX and GNO lower error measurements is a 4.6 sigma effect, which can only occur by chance with probability $2 \times 10^{-6}$. Improvements in the design and operation of the experiments (especially the lower background rate in GNO compared with that in GALLEX) could account for these changes. This difference in the statistical properties of the GALLEX and GNO datasets is a further reason to be cautious about combining them into one dataset.

Since a bimodal flux histogram is suggestive of time variation, it is interesting to examine the flux as a function of time for each experiment. In order to reduce the scatter in the data, we show in Figure 3 the 11-point running mean, and the standard error of the mean, of flux values for the GALLEX and GNO experiments. The two curves look quite different: the GALLEX curve extends from a low of 40 SNU to a maximum of 110 SNU, whereas the GNO curve extends from a low of 40 SNO to a maximum of only 70 SNU.

We next make maximum-likelihood estimates of the flux from each dataset, using the published data. For each run, we have the flux estimate g and their upper and lower error estimates, $\sigma_u$ and $\sigma_l$. For each run, we form a probability distribution function (pdf) for the actual flux, which we denote by f, as follows: the pdf has its peak at g, as is appropriate since g is the maximum-likelihood estimate of the flux; for values of f larger than g, the pdf has the upper half of a normal distribution with width $\sigma_u$; for values of f less than g, the pdf has the lower half of a normal distribution with width $\sigma_l$. We truncate the pdf to be zero for negative values of f, since these are physically prohibited, and then normalize so that the integral of the pdf is unity. In this way, we form, for each experiment, a set of pdf's $P_r(f)$ for runs $r = 1,...,R$, where R is the total number of runs for the experiment.

We now form the likelihood of the flux for each experiment as follows:

$$L(f) = \prod_{r=1}^{R} P_r(f).$$  (3.2)

This quantity (normalized to have maximum value unity) is shown in Figures 4 (a) and (b) for GALLEX and GNO data, respectively. We find from these curves that the maximum likelihood value of the flux is $69.9 \pm 6.1\ SNU$ on the basis of GALLEX data, and $54.1 \pm 5.1\ SNU$ on the basis of GNO data. We may evaluate the difference of the fluxes ($15.8\ SNU$), and the rms error estimate ($7.9\ SNU$), and so find that these two flux estimates differ by 2 sigma, a difference that, for two tails, occurs by chance with probability 0.05.

We note, however, that these flux estimates differ from the recent results of "global" analyses (likelihood analyses of the times of $^{71}Ge$ candidate decay events for all runs considered together) carried out by Cattadori, who obtains the estimates $77.5 \pm 7.7\ SNU$ for GALLEX, and $62.9 \pm 5.4\ SNU$ for GNO [33]. According to these estimates, the flux difference is $14.6\ SNU$ and the rms error estimate is $9.4\ SNU$, so that the difference is not statistically significant. We discuss in Section 7 possible reasons why the flux estimates derived from the summary data for each run may differ from the flux estimates derived by global analyses.

We may carry out a more direct examination of each dataset to look for evidence of variability. Since the number of runs in GALLEX (65) factors into $5 \times 13$, it is convenient to divide the 65 runs of GALLEX data into five blocks of 13 runs each, 1 – 13, 14 – 26, etc., and then use the likelihood method to obtain, for each block, a flux estimate and upper and lower error estimates. The results are given in Table 2. Since the upper and lower error estimates are almost equal, we approximate the real pdf's to normal pdf's with widths given by the means of the upper and lower error estimates. We may then apply the chi-square test [34]. We find the value of the flux (assumed constant) that minimizes the chi-square statistic, which is then found to have the value 12.77. For four degrees of freedom, this can occur by chance with probability 0.012. Hence this test indicates that, at the 99% confidence level, the solar neutrino flux, *as measured by the GALLEX experiment*, is incompatible with the assumption that the solar neutrino flux is constant.

We also divide the GNO data into five blocks, and obtain the flux and error estimates shown in Table 3. We find that the chi-square statistic has the value 4.9. For four degrees of freedom, this can occur by chance with probability 0.29. Hence, according to this test, GNO data are compatible with the assumption that the solar neutrino flux is constant.

## 4. LOMB-SCARGLE POWER-SPECTRUM ANALYSIS

Before beginning power-spectrum analysis, it is important to note that the "sampling" represented by the radiochemical data is neither completely regular nor completely random. We show in Figure 5 the window function (or "periodogram of the sampling")

$$S_T(\nu) = \frac{1}{R} \left| \sum_{r=1}^{R} \exp(i 2\pi \nu t_{er}) \right|^2, \qquad (4.1)$$

formed from the end-times of combined GALLEX-GNO data. We see that the main timing peak is at $\nu_T = 52.18 \, yr^{-1}$, as expected since the run durations are multiples of one week. (Similar periodograms may be formed from GALLEX and GNO datasets, taken separately.) As a result, a real oscillation with frequency $v_0$ will show up also at the alias frequencies $v_T \pm v_0$, $2v_T \pm v_0$, etc. - in particular, at $v_T - v_0$.

We now form a power spectrum of the combined GALLEX and GNO datasets by the Lomb-Scargle procedure [35 - 37]. We first normalize the flux measurements to have mean value zero:

$$x_r = g_r - mean(g). \tag{4.2}$$

We then form the power spectrum from

$$S(v) = \frac{1}{2\sigma_0^2} \left\{ \frac{\left[\sum_r x_r \cos(2\pi v(t_r - \tau))\right]^2}{\left[\sum_r \cos^2(2\pi v(t_r - \tau))\right]} + \frac{\left[\sum_r x_r \sin(2\pi v(t_r - \tau))\right]^2}{\left[\sum_r \sin^2(2\pi v(t_r - \tau))\right]} \right\}, \tag{4.3}$$

where

$$\sigma_0 = std(x), \tag{4.4}$$

and $\tau$ is defined by the relation

$$\tan(4\pi v \tau) = \frac{\sum_r \sin(4\pi v t_r)}{\sum_r \cos(4\pi v t_r)}. \tag{4.5}$$

In order to use the Lomb-Scargle procedure, it is necessary to assign a definite time $t_r$ to each bin. As usual [26], we assign each flux measurement to the end time of each run, since the decay process means that flux measurements are weighted towards the flux value at the end of the run.

This analysis of the combined GALLEX-GNO dataset yields the power spectrum shown in Figure 6. We see that aliasing is very pronounced. The power spectrum is close to mirror-symmetric with respect to the frequency $26\,yr^{-1}$. For the frequency range $0 - 26\,yr^{-1}$, Figure 6 is very similar to Pandola's Figure 4 [26].

5. EXTENDED LOMB-SCARGLE POWER-SPECTRUM ANALYSIS

The Lomb-Scargle procedure is equivalent to fitting the data (normalized to have mean value zero) to a sine wave by the least-squares procedure and evaluating the square of the amplitude at each frequency [19]. We may therefore generalize the procedure for application to measurements made over a finite time interval rather than at points in time.

If we consider that the probability distribution function (pdf) for the flux is a Gaussian centered on g$_r$, with width $\sigma_r$, the likelihood that the data may be fitted to a model that gives G$_r$ as the expected values of g$_r$ is given by

$$P = \prod_{r=1}^{R} \frac{1}{(2\pi)^{1/2} \sigma_r} \exp\left[-\frac{(g_r - G_r)^2}{2\sigma_r^2}\right]. \qquad (5.1)$$

Hence (ignoring an additive constant term) the log-likelihood is given by

$$\Lambda = -\tfrac{1}{2} \sum_{r=1}^{R} (g_r - G_r)^2 / \sigma_r^2. \qquad (5.2)$$

If G$_0$ is the constant value that maximizes the log-likelihood, we write

$$\Lambda_0 = -\tfrac{1}{2} \sum_{r=1}^{R} (g_r - G_0)^2 / \sigma_r^2. \qquad (5.3)$$

Then the power is given [17, 19] by

$$S = \Lambda - \Lambda_0. \qquad (5.4)$$

For values of the frequency $\nu$ defined by the search band and the sampling interval, we consider the model

$$G_r = \frac{k}{\left(1 - e^{-kD_r}\right)} \int_{t_{sr}}^{t_{er}} dt\, e^{-k(t_{er} - t)} \left(K + B e^{i2\pi\nu t} + B^* e^{-i2\pi\nu t}\right), \qquad (5.5)$$

in which the scalar K and the complex parameter B are adjusted to maximize $\Lambda$. In this equation, k $(= 22.15\ yr^{-1})$ is the decay constant of the $^{71}$Ge atoms produced from $^{71}$Ga atoms by neutrino capture, and the duration of each run is given by

$$D_r = t_{er} - t_{sr}. \qquad (5.6)$$

The term preceding the integral converts the number of captures into an equivalent capture rate assumed constant (which is the way that radiochemical data are analyzed).

The Lomb-Scargle power spectrum may be obtained by adopting

$$\sigma_r = \sigma_0 \equiv std(g), \qquad (5.7)$$

and

$$t_{sr} = t_{er} - \delta t, \qquad (5.8)$$

where $\delta t$ is a small quantity. This is equivalent to including a delta function $\delta(t - t_{er})$ in the integral of equation (5.5). In this case, $G_0 = mean(g)$, and we may replace K in equation (5.5) by $G_0$.

We now extend the Lomb-Scargle procedure by retaining equation (5.5) but, instead of adopting (5.8), retain the actual start and stop times in equation (5.5). With this procedure, we generate the power spectra shown in Figures 7, 8, and 9 for GALLEX, GNO, and combined GALLEX-GNO, respectively. In these displays, we show only the frequency range $0-26\,\text{yr}^{-1}$. Any peak above $0-26\,\text{yr}^{-1}$ may be an alias of a peak in the range $26-52\,\text{yr}^{-1}$, but one should also note that any peak in the range $0-26\,\text{yr}^{-1}$ may be an alias of a peak in the range $26-52\,\text{yr}^{-1}$. The top 10 powers for these power spectra are shown in Tables 4, 5 and 6. We see that the principal peaks are more pronounced in these power spectra than they were in the power spectra produced by the simpler procedure of Section 4.

The main goal of this article (as was also the case for Pandola's article [26]) is a search for evidence of rotational modulation. However, whereas Pandola adopts search bands (as wide as $0-26\,\text{yr}^{-1}$) that are not specific to solar rotation, we adopt as our primary search band the specific range $12.5-13.8\,\text{yr}^{-1}$, corresponding to the synodic rotation frequency in an equatorial section of the Sun [38]. [We consider modulations outside this primary search band in the two appendices.] We see that the top two peaks in the GALLEX power spectrum fall in this search band. On the assumption that peaks are randomly placed in frequency, the probability that the top two peaks from power spectra covering the range $0-26\,\text{yr}^{-1}$ would be found in the search band $12.5-13.8\,\text{yr}^{-1}$ is $(1.3/26)^2$, i.e. 0.0025. These peaks do not show up in the GNO power spectrum. The probability that this combination of peaks in the rotation band would show up in *either* the GALLEX or the GNO power spectra is 0.005, i.e. 0.5%.

## 6. MONTE CARLO SIGNIFICANCE TESTS

To obtain robust significance estimates, we need to generate surrogate datasets that resemble the actual data as closely as possible, but are such that the putative signal has been scrambled or otherwise removed. If the distribution of flux values were close to Gaussian, we could use the familiar technique [37] of producing simulated flux values by generating normally distributed random values. However, as we see from Figure 1, the distribution of flux values is nowhere near Gaussian for either GALLEX or GNO. Furthermore, this test rests on the assumption that there is no correlation between the flux estimate and the error estimate, but this assumption is invalid for the GALLEX-GNO datasets. For the radiochemical datasets, there is necessarily a strong correlation between flux estimates and error estimates, since they are both derived from processes that involve Poisson statistics. For the GALLEX and GNO datsets, the correlation coefficient relating the flux estimates and the mean error estimates is 0.67 and 0.71, respectively. [For a purely Poisson process, the mean and the standard deviation are equal, so the fact that they are not equal in these datasets implies that non-Poisson processes are important.]

Another possibility would be to simulate the response of the experiment to a constant neutrino flux by generating hypothetical counts in each run and then computing the estimated flux. However, this would generate a unimodal distribution, so it would not resemble the actual datasets.

Fortunately, there is another option. We can generate Monte Carlo simulations by means of the shuffle process [39]. In this procedure, one divides the data into two subsets, and shuffles one subset. For example, if one were using the Lomb-Scargle procedure [35 - 37] that requires only a flux estimate $g_r$ and a single time estimate $t_r$ for each run, we could randomly reassign the flux estimates to the time estimates. The power-spectrum procedure used in the previous section involves the flux estimates $g_r$ and the two time variables $t_{sr}$ and $t_{er}$. We could consider re-assigning the flux estimates $g_r$ to different pairs $t_{sr}$ and $t_{er}$. However, this poses the problem that the experimental procedure may lead to a correlation between the flux estimates $g_r$ and the durations $D_r$ of the runs (which we should retain). In fact there is very little correlation: for GALLEX data, the correlation coefficient is 0.040, and for GNO data, it is 0.046. Nevertheless, we proceed as follows: We keep the flux values $g_r$ and the durations $D_r$ together, and randomly reassign them among end times $t_{er}$, so retaining the slight correlation between the flux estimates and durations. We are allowed to reassign flux measurements among end times, since we are testing the hypothesis that there is no correlation between these two variables.

We have applied the above procedure to the GALLEX dataset, adopting as a search band the frequency range $12.5 - 13.8 \text{ yr}^{-1}$. For each simulation, we compute the power spectrum over the search band, and note the power of the highest peak in that frequency range, which we denote by SM for "spectral maximum." We then examine the distribution of the maximum-power values. Figure 10 shows the distribution of values of SM from the simulations, and indicates the values of the two prominent peaks in the rotational search band for the actual data listed in Table 4, namely that at 13.64 yr$^{-1}$ with $S = 7.81$ and that at 13.08 yr$^{-1}$ with $S = 6.08$. We find that 55 simulations out of 10,000 have values of SM equal to or larger than that 7.81, and 382 simulations have values of SM equal to or larger than that 6.08. The probability of finding by chance a peak with power 7.81 or more in the rotational search band is 0.55%. The probability of finding by chance both the primary peak and the secondary peak in the rotational search band with powers 7.81 and 6.08 or more is given approximately by the product of the two probabilities, i.e. 0.0002. For GNO, the largest peak in the search band has power only 2.87, which is obviously insignificant, so there is no point in carrying out simulations. Hence there is evidence for rotational modulation in GALLEX data, but no evidence for rotational modulation in GNO data.

## 7. DISCUSSION

Our analyses appear to yield several indications of variability in GALLEX data, but little indication of variability in GNO data. As we discussed in an earlier article [4], and as we noted in Section 3, the histogram of GALLEX flux measurements is clearly bimodal, which (if it is not due to systematic effects) is an indication of variability. Our analysis of subdivisions of GALLEX data also indicates variability, but the same is not true of GNO data.

We also noted in Section 3 that maximum-likelihood analysis of the available data leads to the flux estimates $69.9 \pm 6.1\,SNU$ for GALLEX and $54.1 \pm 5.1\,SNU$ for GNO (which differ by 2 sigma). By comparison, the estimates presented as the result of "global" analyses (i.e. the maximum likelihood analyses of the times of $^{71}Ge$ candidate decay events) are $77.5 \pm 7.7\,SNU$ for GALLEX and $62.9 \pm 5.4\,SNU$ for GNO (which differ by 1.5 sigma) [33]. One should bear in mind that if the flux is variable, as it appears to be from our analyses in Sections 3 and 4, there is no such thing as "the" value of the flux. One could define the mean value, and expect that different analysis procedures will give consistent estimates of this quantity, but the "maximum likelihood" value will depend on the adopted form or forms for the probability distribution function for the flux. Of course, if the data presented to summarize information for each run were a complete representation of that information, analysis of data for all runs should give results identical to those obtained by analysis of the original unbinned data. However, simply giving a best estimate and upper and lower error estimates for each run is clearly an inadequate summary of the relevant experimental information, which is of course the reason that experimentalists undertake the analysis of unbinned data. It is our conjecture, which we plan to discuss in a later article, that, for each run, a complete probability distribution function for all non-negative values of the flux would comprise a complete summary. One could then reasonably expect that the maximum likelihood estimate obtained by combining data from individual runs would agree exactly with the result of a global maximum likelihood analysis.

Our main focus has been upon power spectra formed from GALLEX and GNO data in the search band corresponding to the solar internal rotation rate in an equatorial section of the Sun. This is the proposal originally made in our investigation of Homestake data (except that in the Homestake article [1] we restricted our attention to the radiative zone, whereas we have subsequently adopted a less restrictive search band that includes the convection zone also [7, 17, 18]). For GALLEX data, we see from Figure 7 and Table 4 that there are two peaks in this search band. Monte Carlo simulations carried out in Section 6 indicate that the principal modulation in the rotational search band (at 13.64 $y^{-1}$) would occur by chance with a probability 0.3% (corresponding to a confidence level of 99.7%), and that both modulations (at 13.64 $y^{-1}$ and 13.08 $y^{-1}$) would occur by chance with probability 0.02% (corresponding to a confidence level of 99.98%). There is no indication of rotational modulation in GNO data. As a result, the evidence for rotational modulation is necessarily weaker in a combined GALLEX-GNO dataset than in the GALLEX dataset, as we see from a comparison of Figures 7 and 9.

We also discuss, in Appendices A and B, further evidence for variability. In our power spectrum analysis of GALLEX data, shown in part in Figure 7, we find a peak at 27.32 $y^{-1}$ with power $S = 5.03$. This is very close to twice the primary peak at 13.64 $y^{-1}$. In Appendix A, we carry out a systematic search for evidence of modulation at both a fundamental frequency and the first harmonic of that frequency. We find that there is a probability of only 0.04% that this combination of peaks would occur by chance in the rotational search band, and less than 1% probability that it would occur anywhere in the range $0-26$ $y^{-1}$. In Appendix B, we note that two peaks in the GALLEX power spectrum and one peak in the GNO power spectrum fall in frequency bands that may be associated with r-mode oscillations. Peaks in the rotational search band, in the first harmonic band, and in r-mode bands, are summarized in Table 7.

We now comment briefly on the recent article by Pandola [26], who claimed that his analysis "does not support the presence of a time variability with characteristic periods resembling those of solar magnetic fields." We first note that the strongest peak in Pandola's Figure 1, showing the result of his Lomb-Scargle power-spectrum analysis of the combined GALLEX and GNO I data (84 runs), is at ~ 13.7 $y^{-1}$, which falls in the rotational-modulation search band. This peak also shows up, but with reduced power, in Pandola's Figure 4, that shows the result of his Lomb-Scargle power-spectrum analysis of the combined GALLEX and complete GNO data (123 runs). Although Pandola states in his article that he is looking for evidence of "time variability with characteristic periods resembling those of solar magnetic fields," he in fact adopts for purposes of significance estimation a much wider search band, 0.04 to 26.00 $y^{-1}$, that is not related to solar rotation.

We comment only briefly on Pandola's Figure 7, a power spectrum formed from combined GALLEX-GNO data, since this is somewhat confusing. In terms of the quantity $\lambda$ introduced in Pandola's equation (6), the power is given by $-\ln(\lambda)$. One can recognize in both his Figure 4 and his Figure 7 a peak at 13.55 $y^{-1}$, which we have found to be a strong feature of the GALLEX power spectrum. The confusion arises in connection with the point labeled "best-fit" in Pandola's Figure 7, which does not correspond to a maximum (or to a minimum) in his plot.

The fact that we find evidence of rotational modulation, whereas Pandola does not, may be attributed to the following differences in our analyses: (1) Pandola combines the GALLEX and GNO datasets, whereas we keep them separate. (2) Pandola uses a wide search band that does not focus on rotational modulation, whereas we focus on a comparatively narrow search band that is specific to rotational modulation. (3) We use an extension of the Lomb-Scargle method of power-spectrum analysis that takes account of run duration, whereas Pandola uses the basic Lomb-Scargle procedure that does not take account of run duration. (4) Pandola ignores the possibility that rotational modulation may lead to harmonically related peaks in the power spectrum, but we take account of this possibility.

We defer for a later article power spectrum analysis that takes into account the experimental error estimates. Such an analysis is complicated by several factors: (a) there is a strong correlation between experimental flux estimates and experimental error estimates; (b) the flux histogram for GALLEX data is bimodal; and (c) the error estimates for GALLEX-GNO data are asymmetric and (as we have recently demonstrated [20]), even a slight asymmetry (as in Super-Kamiokande data) can have a notable effect on the resulting power spectrum. It will also be desirable, in future analyses, to take into account the evidence shown in Section 3 that GALLEX data exhibits evidence for variability independent of any rotational modulation.

We acknowledge with thanks helpful communications with Dr Luciano Pandola and support (for PAS) by NSF grant AST-0097128. We also acknowledge helpful discussions with our colleagues A. Kosovichev, J. Pulido, G. Walther, and M.S. Wheatland.

# APPENDIX A. INDICATION OF MODULATION AT THE HARMONIC OF A ROTATION FREQUENCY

Whenever solar activity is modulated by solar rotation, the modulation may occur at a harmonic of the synodic rotation frequency, as well as (or instead of) the fundamental. In our analysis [18] of Super-Kamiokande data, we found that the principal rotational modulation occurs at the second harmonic of the synodic rotation frequency. The Sun-center line-of-sight magnetic field for the same time interval also shows modulation at the second-harmonic but not at the fundamental or the first harmonic. It is therefore interesting to examine GALLEX data for evidence of modulation at either twice or three times $13.64\,yr^{-1}$, the frequency of the principal peak in the rotation band. Three times this frequency is $40.92\,yr^{-1}$ which is high compared with both the decay coefficient $22.15\,yr^{-1}$ and the inverse of the mean run duration ($15.3\,yr^{-1}$), so that it is unlikely to be detectable by the GALLEX experiment. However, we do find a peak at $27.32\,yr^{-1}$ with power 5.03. This frequency is close to $27.28\,yr^{-1}$, the first harmonic of $13.64\,yr^{-1}$.

We may carry out a systematic search for pairs of peaks with frequencies in 2:1 ratio by using the "joint power statistic," introduced recently [39] as a method of looking for correlations in the features of two power spectra. We first form the rms quantity

$$X(\nu) = \left(S(\nu)S(2\nu)\right)^{1/2}, \qquad (A.1)$$

from which the joint power statistic is expressible as

$$J = -\ln\left(2XK_1(2X)\right), \qquad (A.2)$$

where $K_1$ is the Bessel function of the second kind. We find that the following formula gives a close fit to this statistic:

$$J = \frac{1.943 X^2}{0.650 + X}. \qquad (A.3)$$

The important point about this statistic is that if the power spectra are distributed exponentially (which is normally the case), then the joint power statistic is also distributed exponentially.

We have formed the joint power statistic, defined in this way, from the power spectrum shown in Figure 7. The result is shown in Figure 11. We see that, in the band $0-26\,y^{-1}$, the largest peak is found at $\nu = 13.66 \, y^{-1}$ with equivalent power 10.46. There are 14 peaks in the joint power statistic in the rotational search band $12.5 - 13.8 \, y^{-1}$. We may use the "false alarm" formula [35, 36]

$$P = 1 - \left(1 - e^{-J}\right)^N, \qquad (A.4)$$

with $J = 10.46$ and $N = 14$, to infer that there is a probability of only 0.04% of finding such a peak by chance in the rotational search band. For the entire band $0 - 26 \, y^{-1}$, $N = 254$, from which we find that there is a probability of only 0.7% of finding such a peak anywhere in that range.

If we apply the same procedure to the GNO dataset, we find that the greatest value of the joint power statistic over the range $0 - 26 \, y^{-1}$ is 5.17, and the greatest value in the rotational band is only 2.15. These results are not significant.

APPENDIX B. INDICATION OF MODULATION AT R-MODE FREQUENCIES

We now look for evidence of other well known solar oscillations. In our Homestake article [1], we drew attention to the Rieger oscillation [40] (with a period of about 154 days) and related oscillations, which we here refer to as "Rieger-type" oscillations (for instance, oscillations with 72 day period and 51 day period) [41 - 45]. The nature of these oscillations is still a matter of debate [45]. We have proposed [3] that they are due to internal r-mode oscillations [46 - 48]. The frequency of an r-mode oscillation, as seen from Earth, is given by

$$\nu(l, m) = m(\nu_R - 1) - \frac{2 m \nu_R}{l(l+1)}, \qquad (B.1)$$

where $\nu_R$ is the sidereal rotation rate where the oscillation occurs. The interaction of an r-mode with internal (co-rotating) magnetic structures leads to oscillations with frequencies given by

$$\nu(l, m) = \frac{2 m \nu_R}{l(l+1)}. \qquad (B.2)$$

The permissible values of l and m are $l = 2, 3, ...$ and $m = 1, ..., l$. The familiar Rieger-type oscillations given above agree with Equation (B.2) for the values $l = 3$, $m = 1, 2, 3$ and a sidereal rotation frequency of $14.2 \, yr^{-1}$. We have pointed out elsewhere [7, 18] that a prominent oscillation at $9.43\,yr^{-1}$ in Super-Kamiokande data may be attributed to an r-mode oscillation with frequency given by Equation (B.2), for $l=2, m=2$ and a sidereal rotation frequency of $14.15\,yr^{-1}$.

Table 6 includes a list of features in the power spectra formed from GALLEX and from GNO data that may be related to the fundamental and first harmonic of the rotation rate and to low-order r-modes, as given by Equation (B.2), corresponding to $l=2$ and $l=3$. (Note that $\nu(3,2)=\nu(2,1)$, so that these two modes are indistinguishable.) We identify the search bands corresponding to the fundamental and first harmonic of the rotation frequency, and search bands corresponding to these low-order r-mode oscillations frequencies, for sidereal rotation rates in the range $13.50-14.80\,y^{-1}$, corresponding to synodic rotation rates in the range $12.50-13.80\,y^{-1}$. We find in the GALLEX power spectrum weak evidence for modulation that may be associated with the 2-1 or 3-2 mode. We find stronger evidence for modulation that may be associated with the 3-3 mode. In examining the GNO power spectrum, we find evidence for modulation that may be associated with the 2-2 mode (the same mode that shows up in our analysis of Super-Kamiokande data [7, 18]). In the Super-Kamiokande analysis, this modulation appears with frequency $9.43\,y^{-1}$, whereas in our current analysis of GALLEX-GNO data, the frequency is $9.20\,y^{-1}$. This would pose a problem if we were to assume that the r-mode oscillation is stable in frequency, long lived, and has the same effect on the high-energy neutrinos detected by Super-Kamiokande and the low-energy neutrinos detected by GALLEX-GNO. However, Rieger-type oscillations [40 – 45] (which we interpret as r-mode oscillations) are typically transient and their frequencies typically drift. Also, because of the difference in neutrino energies, the two experiments would have oscillation resonances at different radii, so that the modulation frequencies (that are set by the rotation rates) will be different. For all of these reasons, the frequency discrepancy does not in itself rule out the interpretation of the peak at $9.20\,y^{-1}$ as an r-mode oscillation.

It is interesting to note that the expected frequency of the 3-2 r-mode, as given by (B.1), is $20.50-22.67\,y^{-1}$. Hence it is possible that the prominent feature of the GNO power spectrum at $21.93\,y^{-1}$ is also a manifestation of an r-mode oscillation.

TABLE 1

GALLEX Data.

| Runs | Flux | Upper Error | Lower Error |
|---|---|---|---|
| 1 - 13 | 78.6 | 18.3 | 18.1 |
| 14 – 26 | 79.2 | 14.0 | 13.9 |
| 27 – 39 | 63.5 | 12.3 | 11.9 |
| 40 – 52 | 44.3 | 10.6 | 10.1 |
| 53 - 65 | 108.5 | 16.4 | 15.6 |

TABLE 2

GNO Data.

| Runs | Flux | Upper Error | Lower Error |
|---|---|---|---|
| 1 - 12 | 51.3 | 12.2 | 11.9 |
| 13 – 23 | 58.2 | 12.5 | 12.1 |
| 24 – 35 | 72.3 | 12.3 | 11.9 |
| 36 – 46 | 55.4 | 11.8 | 11.7 |
| 47 - 58 | 38.5 | 9.9 | 9.8 |

TABLE 3

Top ten peaks in the GALLEX power spectrum over the frequency range 0 to $26 yr^{-1}$.

| Frequency ($yr^{-1}$) | Power |
|---|---|
| 13.64 | 7.81 |
| 13.08 | 6.08 |
| 4.54 | 5.73 |
| 11.87 | 5.51 |
| 6.93 | 5.02 |
| 21.55 | 4.87 |
| 6.05 | 4.81 |
| 3.97 | 4.62 |
| 25.59 | 4.17 |
| 12.26 | 3.83 |

TABLE 4

Top ten peaks in the GNO power spectrum over the frequency range 0 to $26 yr^{-1}$.

| Frequency ($yr^{-1}$) | Power |
|---|---|
| 21.93 | 5.45 |
| 2.93 | 4.80 |
| 3.52 | 4.25 |
| 9.20 | 4.10 |
| 17.29 | 4.00 |
| 16.53 | 3.70 |
| 15.92 | 3.54 |
| 10.07 | 3.30 |
| 1.10 | 3.18 |
| 23.03 | 2.98 |

TABLE 5

Top ten peaks in the GALLEX-GNO power spectrum over the frequency range 0 to $26\,yr^{-1}$.

| Frequency (yr$^{-1}$) | Power |
|---|---|
| 11.82 | 6.32 |
| 13.63 | 5.44 |
| 2.94 | 5.36 |
| 23.33 | 5.28 |
| 10.72 | 5.08 |
| 6.37 | 4.64 |
| 10.08 | 4.62 |
| 23.07 | 4.59 |
| 6.95 | 4.42 |
| 6.07 | 4.29 |

Table 6

Peaks in GALLEX and GNO power spectra.

|  |  | GALLEX Frequency | GALLEX Power | GNO Frequency | GNO Power |
|---|---|---|---|---|---|
| Rotation, fundamental | 12.50 to 13.80 | 13.08 | 6.08 |  |  |
|  |  | 13.64 | 7.81 |  |  |
| Rotation, harmonic | 25.00 to 27.60 | 27.32 | 5.03 |  |  |
| R-modes, 2-1 and 3-2 | 4.50 to 4.93 | 4.54 | 5.73 |  |  |
| R-mode, 2-2 | 9.04 to 9.87 |  |  | 9.20 | 4.10 |
| R-mode, 3-1 | 2.25 to 2.46 |  |  |  |  |
| R-mode, 3-3 | 6.75 to 7.40 | 6.93 | 5.02 |  |  |

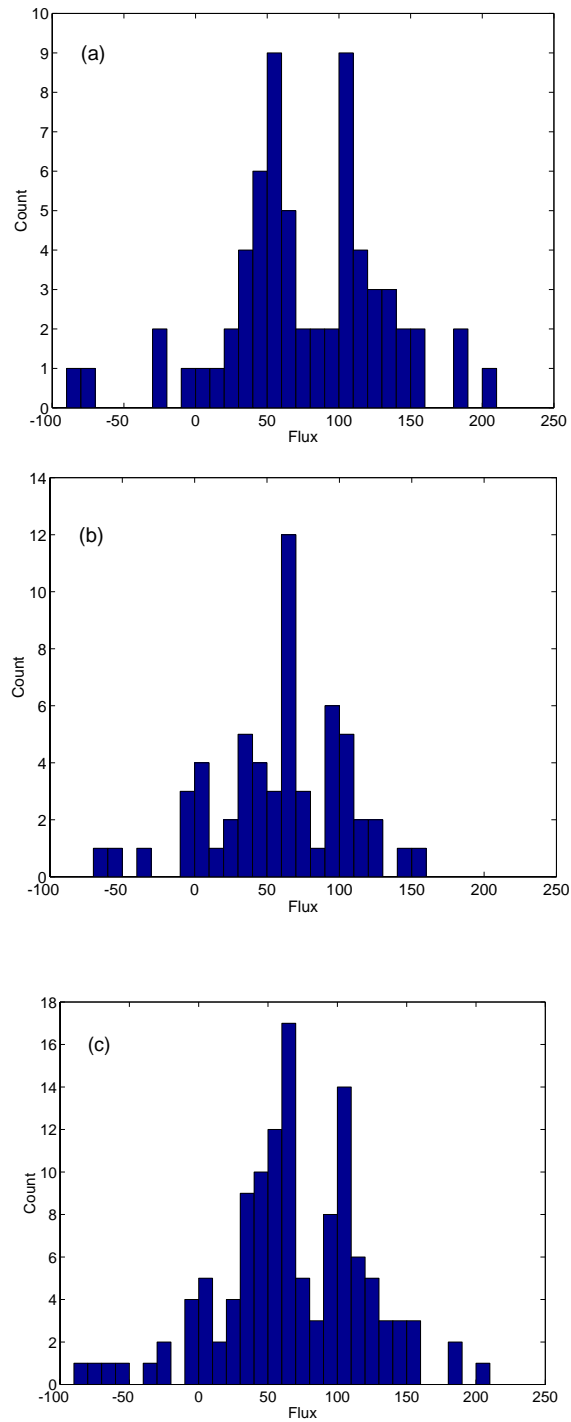

Figure 1. Histograms of flux values for (a) GALLEX, (b) GNO, and (c) combined GALLEX-GNO.

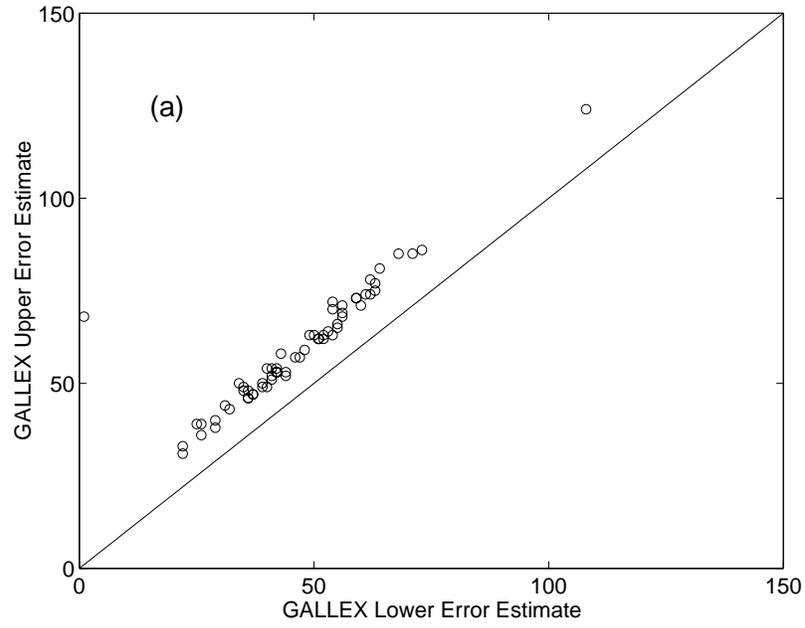

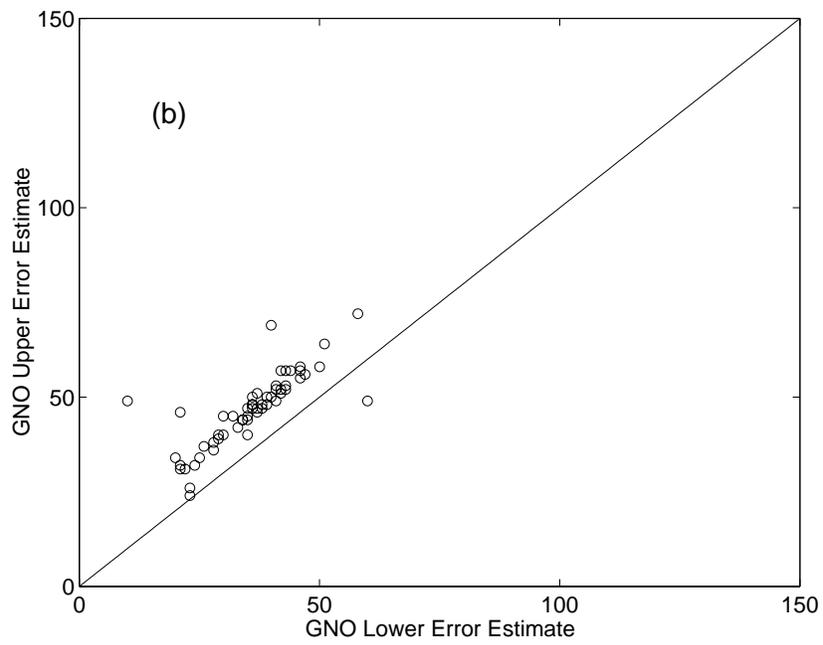

Figure 2. Plots of error estimates for (a) GALLEX and (b) GNO.

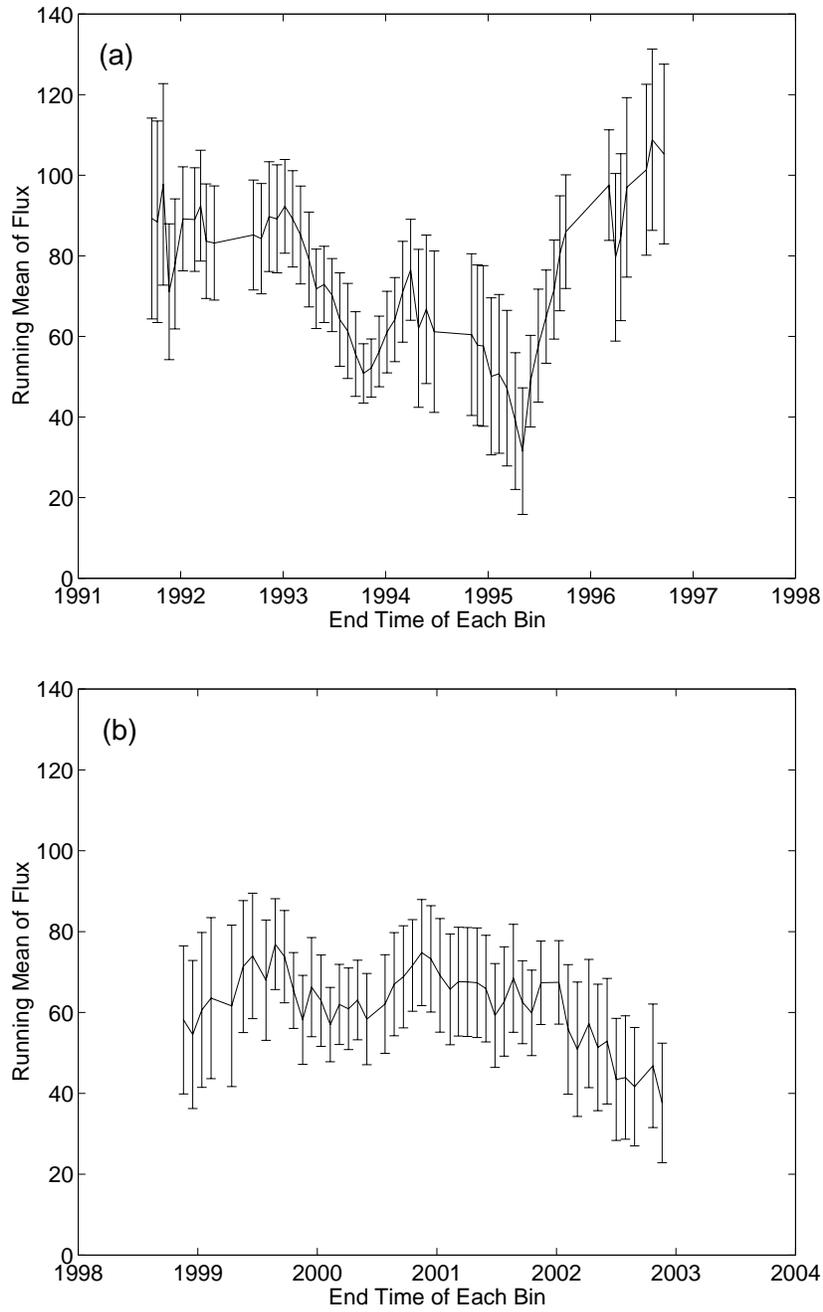

Figure 3. Plots of 11-point running mean and standard error of the mean for the experimental flux estimates for (a) GALLEX and (b) GNO.

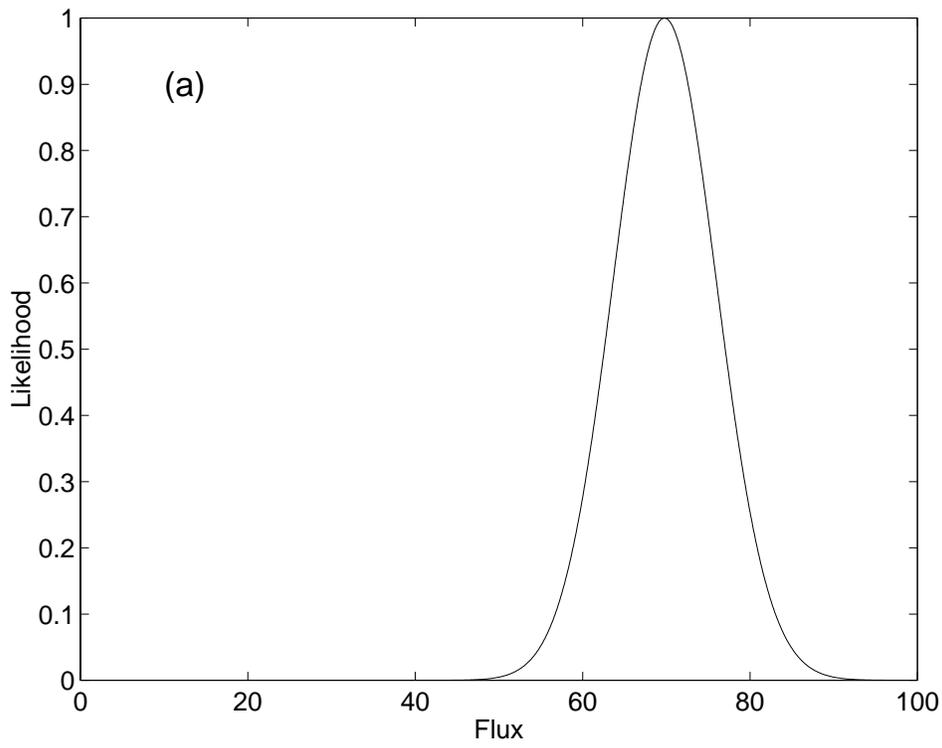

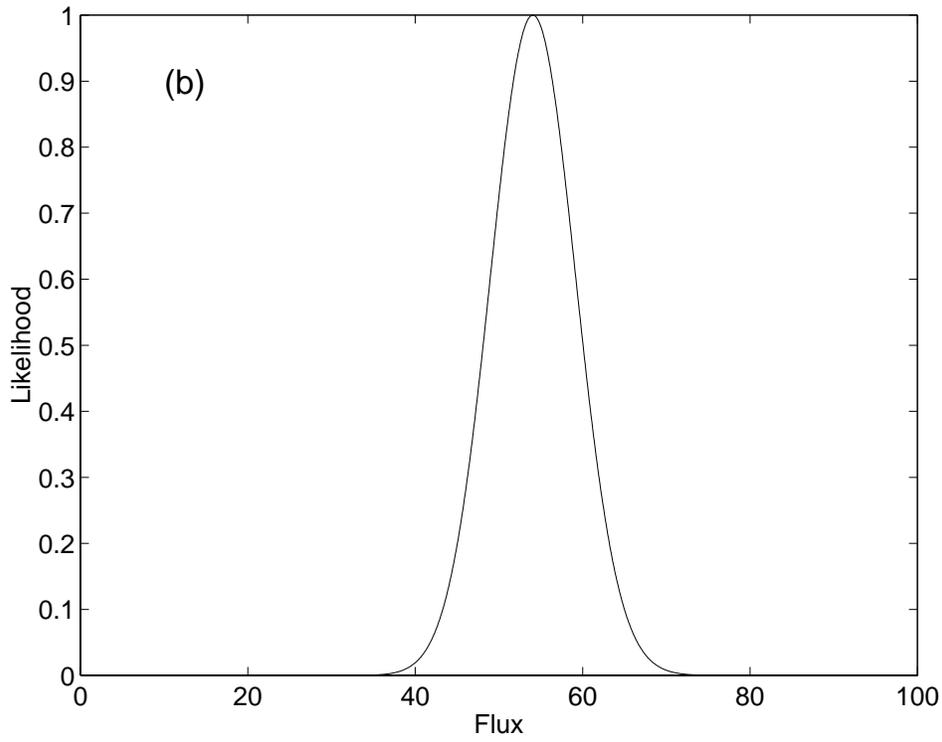

Figure 4. Likelihood function for the flux for (a) GALLEX and (b) GNO.

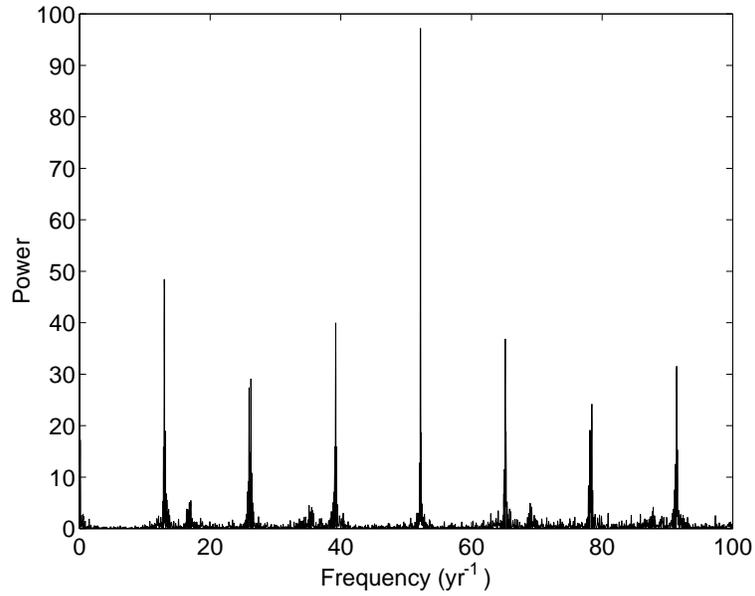

Figure 5. Periodogram computed from end-times of GALLEX-GNO data.

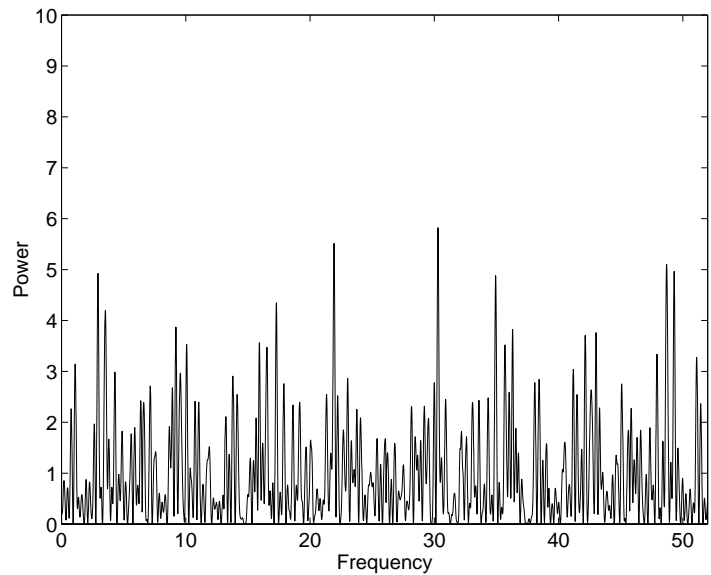

Figure 6. Power spectrum derived, by the Lomb-Scargle procedure, from combined GALLEX-GNO data for the frequency range $0-52\ yr^{-1}$. Note the near-symmetry with respect to the frequency $26\ yr^{-1}$.

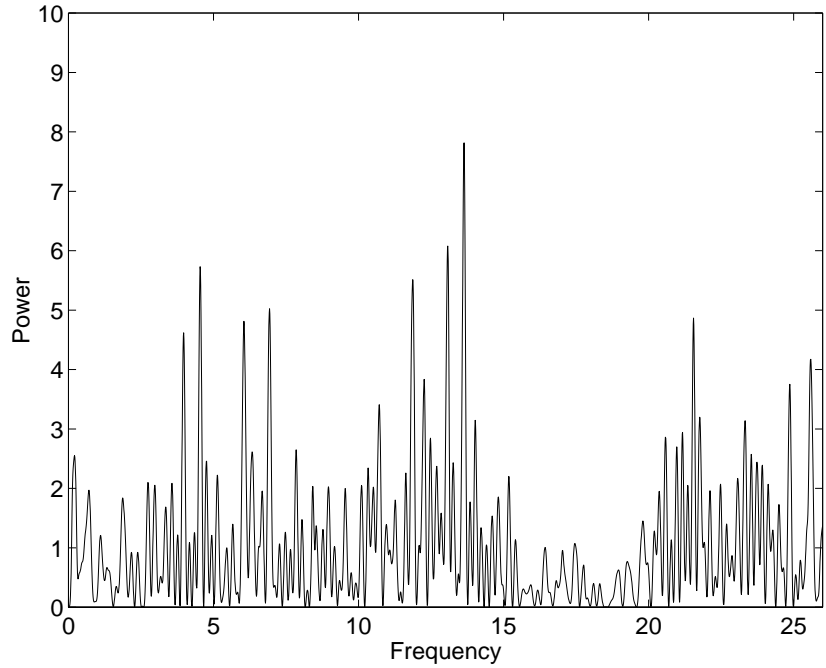

Figure 7. Power spectrum derived, by an extension of the Lomb-Scargle procedure, from GALLEX data for the frequency range $0-26\ yr^{-1}$.

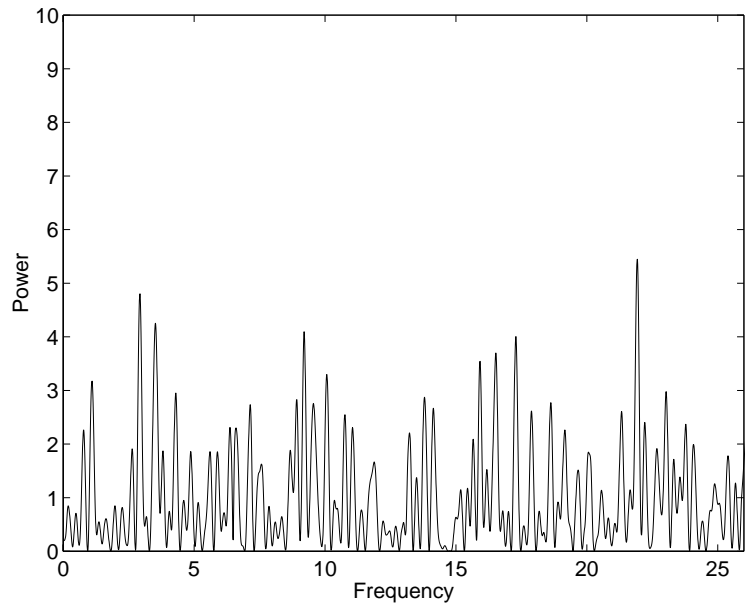

Figure 8. Power spectrum derived, by an extension of the Lomb-Scargle procedure, from GNO data for the frequency range $0-26\ yr^{-1}$.

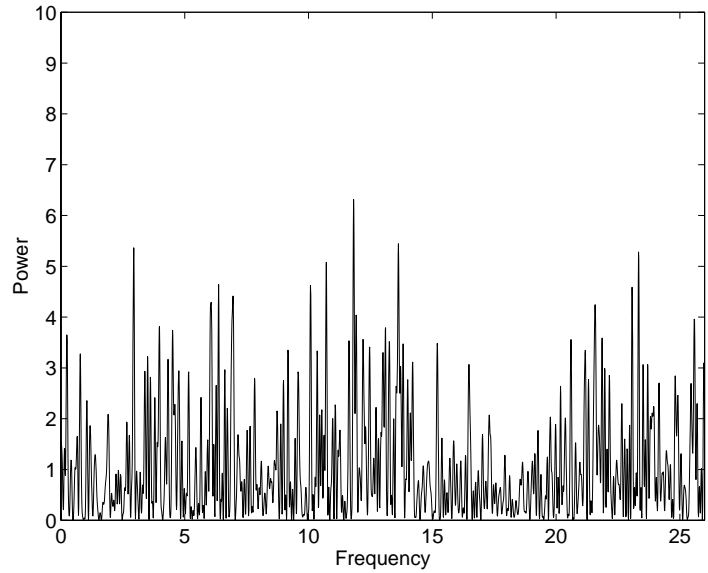

Figure 9. Power spectrum derived, by an extension of the Lomb-Scargle procedure, from combined GALLEX-GNO data for the frequency range $0-26\ yr^{-1}$.

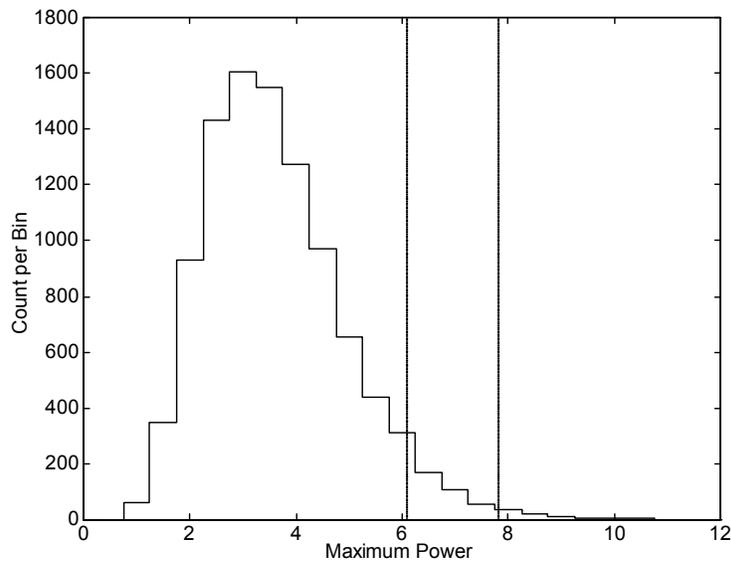

Figure 10. Histogram of maximum power in the rotational search band $12.5-13.8\ yr^{-1}$ for 10,000 simulations of GALLEX data. The vertical lines indicate the power of the primary (7.81) and secondary (6.08) peaks in the search band.

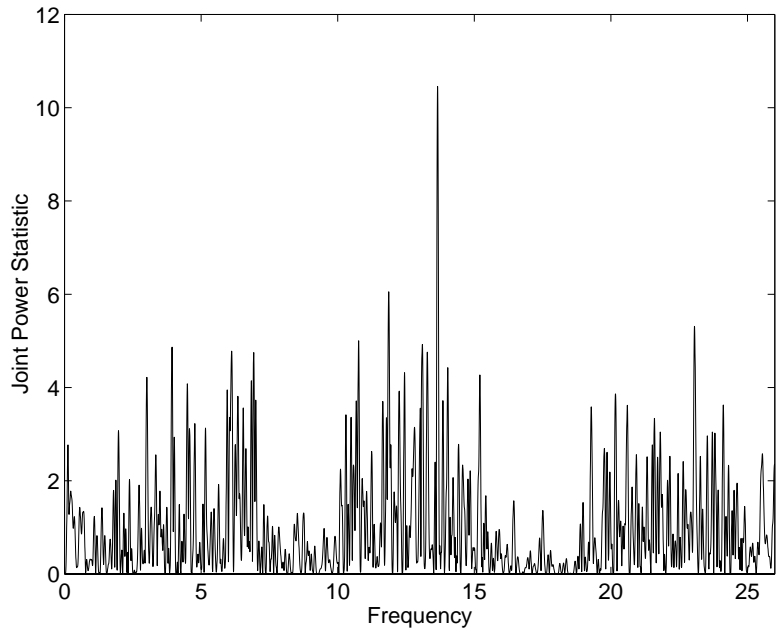

Figure 11. Joint power statistic formed from power at $\nu$ and at $2\nu$ for GALLEX data.